# Fractal Texture and Structure of Central Place Systems

Yanguang Chen

(Department of Geography, College of Urban and Environmental Sciences, Peking University, Beijing 100871, P.R. China. Email: chenyg@pku.edu.cn)

**Abstract**: The boundaries of central place models proved to be fractal lines, which compose fractal texture of central place networks. A textural fractal can be employed to explain the scale-free property of regional boundaries such as border lines, but it cannot be directly applied to spatial structure of real human settlement systems. To solve this problem, this paper is devoted to deriving structural fractals of central place models from the textural fractals. The method is theoretical deduction based on the dimension rules of fractal sets. The textural fractals of central place models are reconstructed, the structural dimensions are derived from the textural dimensions, and the central place fractals are formulated by the *k* numbers and *g* numbers. Three structural fractal models are constructed for central place systems according to the corresponding fractal dimensions. A theoretical finding is that the classic central place models comprise Koch snowflake curve and Sierpinski space filling curve, and an inference is that the traffic principle plays a leading role in urban and rural evolution. The conclusion can be reached that the textural fractal dimensions can be converted into the structural fractal dimensions. The latter dimensions can be directly used to appraise urban and rural settlement distributions in the real world. Thus, the textural fractals can be indirectly utilized to explain the development of the systems of human settlements.

**Key words**: Central place fractals; Systems of human settlement; Fractal dimension; Koch snowflake; Sierpinski space-filling curve; Gosper island

# 1. Introduction

The world we live in is actually a fractal world rather than a Euclidean world, because the urban and rural settlements are organized by fractal principles. Central place theory is one of the most important classic doctrines on urban settlement systems, concerned with the urban and rural organization that associates spatial network with hierarchical order. The theoretical models are presented by Christaller (1933/66) and underpinned by Lösch (1940), and developed by many geographers. The theory reveals the way that human settlement systems grow, evolve, and are spaced out, and lay the foundation of development of human geography (Longley *et al*, 1991). Thirty years ago, an excellent study was made by Arlinghaus (1985), who brought to light the fractal texture from central place models (Arlinghaus, 1993). This work made a precedent for the research on central place fractals. Later, the relationships between fractal dimension and arbitrary Löschian numbers are revealed by means of Diophantine analysis (Arlinghaus and Arlinghaus, 1989). Arlinghaus' studies show us a new way of deep understanding the growth and evolution of human settlement systems. What is more, geographical fractal lines such as regional boundaries can be interpreted by analogy with the textural central place fractals. As we know, national boundary lines bear fractal property and once puzzled many people (Mandelbrot, 1982). One of the reasons rest with that the processes of human settlement growth influence the patterns of regional boundaries.

In the last 20 years, significant progress has been made in fractal city research. In particular, multifractal theory and method have been introduced into urban studies, including the research on central place systems (Ariza-Villaverde *et al*, 2013; Chen, 2014; Chen and Wang, 2013; Hu *et al*, 2012; Murcio *et al*, 2012). Unfortunately, a paradox arose because of the contradiction between the theoretical prediction and observational evidences of central place studies. Central place models are based on the triangular lattice patterns of urban places and regular hexagonal tessellations of complementary areas. One of its basic assumptions is perfect space filling. Despite the hierarchical difference, the spatial patterns of central places indicate a homogeneous distribution with a Euclidean dimension $d$=2. On the other hand, according to the observational data of urban settlements in southern Germany (Christaller, 1933/66), the spatial distributions take on the fractal characters indicating heterogeneous distribution and the fractal dimension values of central place networks come between 1.4 and 1.9 (Chen and Zhou, 2006). These values suggest fractal structure of central place systems. Allen and his co-workers used the dissipative structure theory to explain the symmetry breaking and spatial heterogeneity of central place systems (Allen, 1982; Allen, 1997; Prigogine and Stengers, 1984). Chen employed the ideas from intermittency to explain the fractal structure of central place system (Chen, 2011), and adopted multifractal scaling to describe the spatial heterogeneity of settlement distributions (Chen, 2014). Thus the theories of fractals and self-organization can be used to eliminate the paradox of space dimension of central place systems.

However, the relationships between the fractal texture and structure of central place systems are not yet clear. A central place network contains points (nodes), lines (sides), and area (service region). The lines indicate spatial texture of a central place system, while the combination of points with lines suggests the spatial structure of the central place system. The relations between the points, lines, and areas follow allometric scaling laws. Arlinghaus's works illustrate the fractal texture of central place models (Arlinghaus, 1985; Arlinghaus, 1993), while Chen's studies demonstrate the fractal structure of real central place systems (Chen, 2011; Chen and Zhou, 2006). What is the theoretical relation between the textural fractals and structural fractals? It is easy to describe real systems of human settlements by structural fractal models of central place system, but the textural fractal models cannot be directly applied to urban and rural places. This paper is devoted to exploring the logic connection between the fractal texture and structure. The specific research objectives are as below: (1) Derive the mathematical relation between the textural fractal dimension and structural fractal dimension; (2) Construct the fractal models for the spatial structure of central place systems in light of the structural fractal dimensions; (3) Present a set of examples to illustrate how to use the new fractal parameter equations. An unexpected discovery is that the traffic principle plays a leading role in the evolution of the real central place systems. The rest parts of the paper are organized as follows. In Section 2, the latent fractals in central place models are revealed, and new fractal models are proposed. A set of fractal dimension formulae are presented for fractal texture and structure of central place systems. In Section 3, an empirical analysis is made for real systems of urban places, and the results inspire new understanding about central place principles. In Section 4, several related questions are discussed. Finally, the discussion is concluded with summarizing the main points of this study.

## 2. Theoretical results

**2.1 Fractal texture of central place systems**

The spatial patterns of urban settlements are organized by self-similar principle and human fractals are previous to fractal theory. A number of regular fractals are hidden behind the hierarchies of hexangular cells in central place networks. Among these fractals, three types of fractal shapes have been brought to light by Arlinghaus (1985). This suggests that central place fractals can be employed to define the boundary of urban service area. To characterize central place fractals, two numbers can be defined to re-express central place fractal dimension. One is the *k* number, and the other, *g* number. The *k* number is in essence the common ratio of the geometric sequences of a central place hierarchy. In literature, the *k* numbers are termed Löschian numbers, in which the three special ones are Christaller's number (3, 4, 7)[1]. A Diophantine analysis can be employed to create arbitrary Löschian numbers (Arlinghaus and Arlinghaus, 1989). The *g* number is defined by the amount of fractal units in a fractal generator. For example, for the Koch curve, there are 4 line segments in the fractal generator (Beck and Schlögl, 1993; Mandelbrot, 1982). Thus, the *g* number is 4. For the Sierpinski gasket, there are 3 triangles in the fractal generator (Beck and Schlögl, 1993; Mandelbrot, 1982). Thus, the *g* number is 3. The *g* number can be also termed similarity point number. Therefore, a Koch curve has 4 similarity points, a Sierpinski gasket has 3 similarity points, and so on. The rest cases may be deduced by analogy.

The textural fractal dimension of central place models can be given by the *k* and *g* numbers. The formula is $D_T = 2 \ln(g)/\ln(k)$, where $D_T$ denotes the textural fractal dimension. The fractal dimension values of three types of central place models are as below: For $k=3$ system, $g=2$, and thus the fractal dimension $D_T = \ln(2)/\ln(3^{1/2}) = 1.2619$; For $k=4$ system, $g=3$, and the fractal dimension $D_T = \ln(3)/\ln(2) = 1.5850$; For $k=7$ system, $g=3$, and the fractal dimension $D_T = \ln(3)/\ln(7^{1/2}) = 1.1292$. These fractal dimension values have been illustrated by Arlinghaus (Arlinghaus, 1985; Arlinghaus, 1993). In fact, more fractals can be identified from the models of central places. In this section, I will demonstrate that the three types of central place models contain Koch snowflake curve, Sierpinski space-filling curve, and Gosper curve (Appendix 1). These fractals were partially discussed by Mandelbrot (1982). All these curves reflect fractal lines, and a self-similar hierarchy of scale-free line segments forms the fractal texture. The spatial elements of textural pattern are mainly fractal lines.

The first type of central place systems is marked by *k*=3, which indicates the marketing principle.

---

[1] The *k* number is first defined by Christaller (1933) and developed by Lösch (1940). Christaller (1933) only gave 3 three *k* numbers, that is, 3, 4, and 7, which can be termed Christaller's numbers. The generalized Christaller's numbers are called Löschian number.

Where shape is concerned, the service area of a $k=3$ central place is in essence equivalent to a Koch island. On the one hand, clearly, the textural fractal dimension of $k=3$ system is equal to the fractal dimension of Koch curve, $D=\ln(4)/\ln(3)=1.2619$. On the other hand, we can create a Koch snowflake curve using the generator of the $k=3$ central place fractal. For the fractal boundary of the $k=3$ system, the generator comprises two line segments, between which there is an included angle of 120 degrees (Arlinghaus, 1985; Arlinghaus, 1993) (Appendix 1). As indicated above, the number of line segments in the generator can be termed $g$ number ($g=2$) (Table 1). The length of each line segment in the generator is $1/3^{1/2}$ of the length of each line segment in the initiator. We have two ways of applying the generator to the initiator to construct fractal lines. One is to change the direction of generator by segments, and the other is to change direction by steps. Suppose that the initiator is a regular hexagon. Applying the generator to the initiator inward and outward in turns by segments yields the second polygon, and the length of each new line segment is reduced to $1/3^{1/2}$ of the original line segment. Then applying the generator to the second polygon inward and outward by segments yields the third polygon, and the length of each new line segment is reduced to $1/3$ of the original line segment. The rest can be done in the same manner. Finally, we can produce a fractal boundary of central place service area with dimension $D=1.2619$ (Figure 1(a)). Using the same initiator and generator, we can construct a Koch snowflake curve. Applying the generator to the initiator inward collectively yields a hexangular star, then applying the generator to the hexagram outward and outward yields a hexapetalous curve (Chen, 1998a). The rest may be treated by analogy. Finally, we can derive a Koch island (Figure 1(b)). The Koch snowflake curve does not represent the service area of a central place, but it reflects the growing pattern of a central place network.

The second type of central place systems is marked by $k=4$, which indicates the traffic principle. The fractal boundary of a $k=4$ central place is in fact a Sierpinski space-filling curve. On the one hand, the fractal dimension of $k=4$ system is equal to that of Sierpinski gasket. On the other, we can derive a hybrid Sierpinski space-filling curve from the $k=4$ central place network. For the fractal boundary of the $k=4$ system, the generator comprises three line segments ($g=3$), between which there are two included angles of 120 degrees (Arlinghaus, 1985; Arlinghaus, 1993) (Appendix 1). The length of each line segment in the generator is $1/2$ of the line segment length in the initiator. There are two approaches to constructing the $k=4$ central place fractal lines. One is to follow the intensionality principle, and the other is to follow the extensionality principle. Arlinghaus (1985) created the *k*=4 fractal texture according to the intension rule. In this paper, I will derive the fractal boundary by means of the extension rule. The initiator is still a regular hexagon. Introducing the generator to the initiator and to the new line segments again and again yield a fractal filling curve (Figure 2(a)). The result corresponds to a mixed Sierpinski gasket, which comprises three overlapping combination Sierpinski gaskets rather than a single one (Figure 2(b)). There are at least two variants of Sierpinski gasket. One is extended Sierpinski gasket, and the other is mixed Sierpinski gasket (Appendix 2). Thus the fractal curve can be termed hybrid Sierpinski space-filling curve. According to the fractal combination principle, the fractal dimension of the hybrid Sierpinski gasket is still $D=\ln(3)/\ln(2)=1.5850$. In practice, the random Sierpinski gasket can be employed to model the *k*=4 central place network (Chen, 1998b).

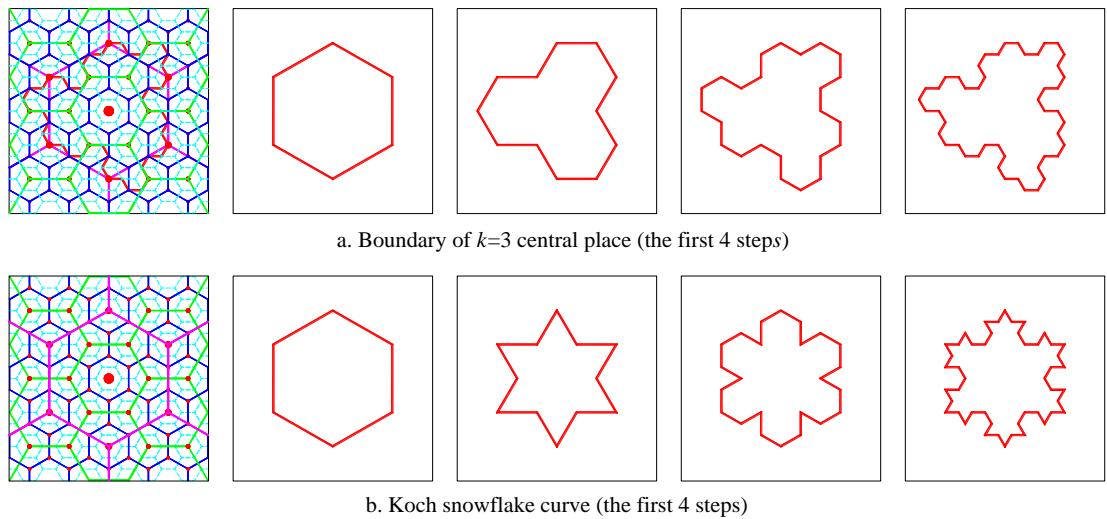

a. Boundary of *k*=3 central place (the first 4 step*s*)

b. Koch snowflake curve (the first 4 steps)

**Figure 1 The central place fractal for the *k*=3 system and the Koch snowflake curve (the first four steps)**

The third type of central place systems is marked by *k*=7, which indicates the separation principle. The fractal boundary of a *k*=7 central place forms the Gosper island. The fractal dimension of the *k*=7 system is equal to that of Gosper curve, $D=\ln(3)/\ln(7^{1/2})=1.1292$ (Mandelbrot, 1982). For the *k*=7 system, the generator comprises three line segments (*g*=3), between which there are two included angles of 120 degrees (Arlinghaus, 1985; Arlinghaus, 1993) (Appendix 1). The length of each line segment in the generator is $1/7^{1/2}$ of the line segment length in the initiator. Applying the generator to the initiator and to the new line segments repeatedly yield a variant of Gosper's island (Figure 2(a)). If we show the hierarchy of boundaries, the patterns is similar to the Gosper flake (Gardner, 1976). Mandelbrot (1982) used this fractal curve to represent rivers and watersheds. In fact, there is an analogy between central place network and systems of rivers associated with watershed area (Woldenberg and Berry, 1967). The $k$=7 central place fractals can be employed to explain the scale-free national boundaries.

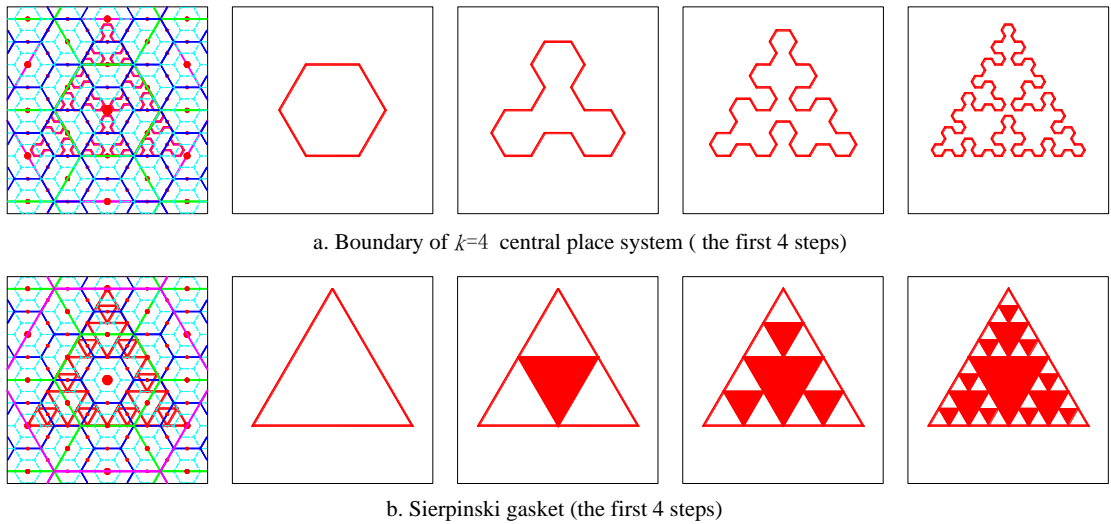

a. Boundary of $k$=4 central place system (the first 4 steps)

b. Sierpinski gasket (the first 4 steps)

**Figure 2 The central place fractal for the $k$=4 system and the Sierpinski gasket (the first four steps)**

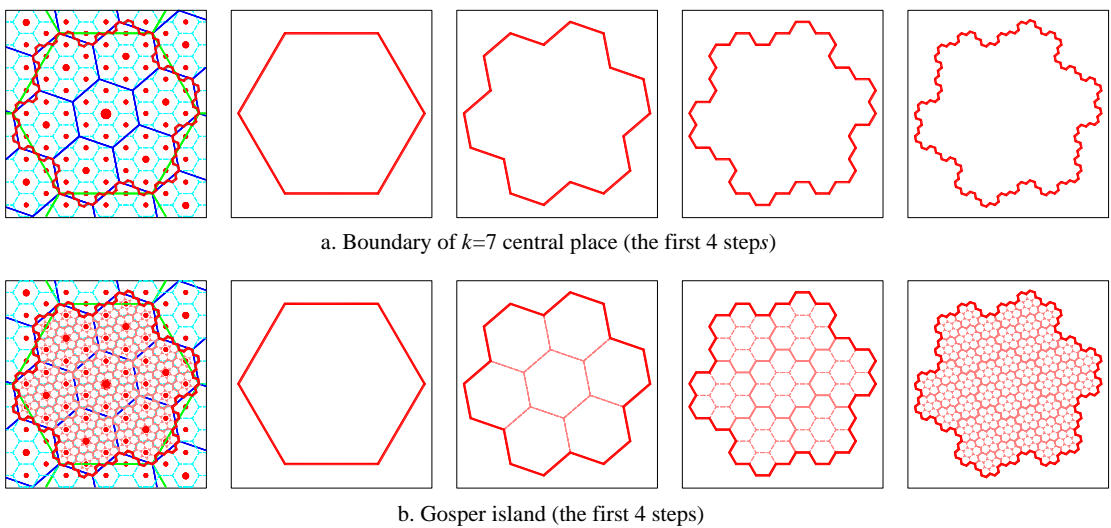

a. Boundary of $k$=7 central place (the first 4 steps)

b. Gosper island (the first 4 steps)

**Figure 3 The central place fractal for the $k$=7 system and the Gosper island (the first four steps)**

Two fractals are worth explaining and emphasizing here because they are helpful for us to understand central place systems. One is the Koch snowflake curve, and the other is the Sierpinski space-filling curve. In the theoretical models, central places compose triangular lattice, and the service regions of all the central places in a given level constitute hexagonal tessellations. If a system of central places develops from a center and its growth is isotropic, the central place system will form a Koch island (Figure 4). The fractal dimension of the Koch curve is 1.2619 (Mandelbrot, 1982). This curve helps us understand the way by which the systems of cities and towns spread from the core to the periphery (Chen, 1998a). As indicated above, for the $k$=4 system, a set of Pascal's triangles can be found, and these Pascal's triangles form a hybrid Sierpinski space-filling curve. If the curve develops and develops, the final result will filling a mixed Sierpinski gasket, which comprises three overlapping Sierpinski gaskets (Figure 5). The fractal dimension of the space-filling curve is 1.5850 (Mandelbrot, 1982). This curve helps us understand the way by which the transport networks are organized in a heterogeneous space (Chen, 1998b).

**Table 1 The textural fractal dimensions and the corresponding fractal equivalences of three central place models**

| Type | Principle | $g$ number | Equivalence of textural fractal |
|---|---|---|---|
| **The $k$=3 system** | Marketing principle | 2 line segments | Koch snowflake curve |
| **The $k$=4 system** | Traffic principle | 3 line segments | Sierpinski space-filling curve |
| **The $k$=7 system** | Separation principle | 3 line segments | Gosper island |

**Note**: In literature, the "market principle" is also termed "marketing principle" or "supplying principle"; "traffic principle" is also termed "transportation principle", and the "separation principle" is also termed "administration principle".

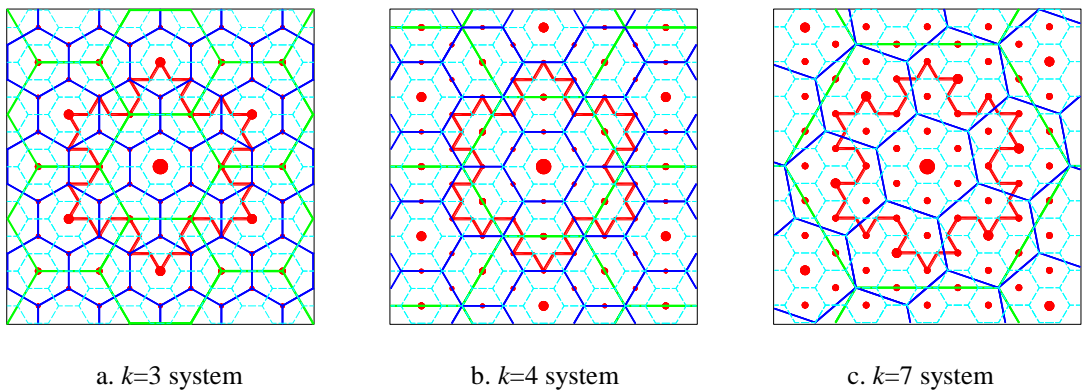

a. $k$=3 system      b. $k$=4 system      c. $k$=7 system

**Figure 4 The Koch snowflake curves in three types of central place networks (the fourth step)**

(**Note**: From the smaller to the larger, we can draw many Koch snowflake curves.)

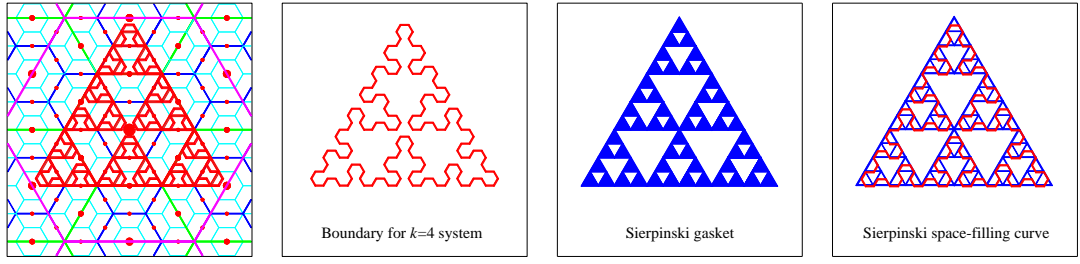

Boundary for *k*=4 system     Sierpinski gasket     Sierpinski space-filling curve

**Figure 5 The correspondence of the hybrid Sierpinski space-filling curve of the *k*=4 system to the mixed Sierpinski gasket (the fourth step)** (**Note**: In a *k*=4 network model of central places, we can find a number of hybrid Sierpinski space-filling curves which are weaved with each other.)

## 2.2 Fractal structure of central place systems

The textural fractal dimension of central place models can be utilized to predict the structural fractal dimension of the central place networks. Fractal texture is mainly reflected by scale-free distribution of lines, while fractal structure can be reflected by the scale-free distribution of points, lines, and areas. The classical central place theory suggests an integral dimension of urban space. In this instance, we can only find fractal texture. No fractal structure in a strict sense can be found. As indicated above, textural fractal models cannot be directly applied to real systems of human settlements. Fortunately, using the ideas from intermittency, we can reveal the fractal structure of central place systems by reference to the fractal texture in central place models. The fractal structure models can be directly applied to real settlement systems. To understand the structure of central place fractals, we should know the hierarchical scaling. Two basic laws of central place systems have been brought to light by theoretical and empirical studies: one is hierarchical scaling law, and the other is spatial coordination law (Chen, 2014). The hierarchical scaling laws can be abstracted as a set of exponential functions or a set of power laws. Suppose that the first order central place serves one hexagonal region. The area of the hexagonal region equals the total area of *k* hexagonal regions of the second order, or the total area of $k^2$ hexagonal regions of the third order. Generally speaking, for the *m*th order, the number of hexagonal regions is as follows

$$N_m = k^{m-1}, \tag{1}$$

where $N_m$ denotes the number of service areas (*m*=1,2,3,…). That is to say, a top-level service area is equivalent to $k^{m-1}$ areas at the *m*th level. The number of central places in the *m*th level is just the difference of the number of complementary regions, that is

$$f_m = \Delta N_m = N_m - N_{m-1} = k^{m-1} - k^{m-2} = (1-\frac{1}{k})k^{m-1}, \qquad (2)$$

in which $f_m$ represents the central place number. Thus we have

$$f_m = f_1 k^{m-1}, \qquad (3)$$

where the coefficient $f_1=1-1/k$. The distance between two adjacent central places at the same level can be expressed as

$$L_m = L_1(\sqrt{k})^{1-m}, \qquad (4)$$

where $L_m$ refers to the distance, and $L_1$ denotes the coefficient indicating the distance between the first order central places. According to equations (3) and (4), the $k$ number is a central place number ratio and squared distance ratio, namely

$$k = \frac{f_m}{f_{m-1}} = (\frac{L_{m-1}}{L_m})^2. \qquad (5)$$

From equations (3) and (4) it follows an inverse square law such as

$$f_m = \mu L_m^{-d}, \qquad (6)$$

where $\mu=f_1 L_1^d$ is proportionality coefficient, and $d=\ln k/\ln\sqrt{k}=2$ is scaling exponent. This suggests that the space dimension of standard central place models is of Euclidean dimension. However, this theoretical prediction is in conflict with the results from the central place systems in southern Germany. According to the observation datasets of Christaller (1933/66), the space dimension of central place systems in Southern Germany come between 1.4 and 1.9, which are fractional dimension instead of Euclidean dimension (Chen and Zhou, 2006). By introducing intermittency process of space filling into urban evolution, we can obtain structural fractals of central place systems (Chen, 2011). In that case, Euclidean dimension is replaced by a fractal dimension, and equation (6) will be replaced by the following relation

$$f_m = \mu L_m^{-D}, \qquad (7)$$

where $D<d=2$ refers to the structural dimension of central place fractals. Accordingly, the inverse square law, equation (5), will be substituted by a power law as follows

$$k = \frac{f_m}{f_{m-1}} = (\frac{L_{m-1}}{L_m})^D. \qquad (8)$$

Compared with equation (5), the distance ratio becomes larger. The structural fractal models can be used to describe the real systems of urban and rural places.

## 2.3 Spatial structure models of central place systems

The dimension relations between fractal texture and fractal structure of central place systems can be derived from fractal dimension principles. A number of fractal dimension rules can be found in literature, including the intersection rule and union rule of fractal sets (Vicsek, 1989, page 17). Using the fractal dimension principles, we can derive the structural dimensions from the textural dimensions of central place fractals. As indicated above, the textural dimensions of fractal central place systems have been revealed by Arlinghaus (1985). Further, by means of the structural fractal dimension values, we can derive more textural fractal dimension values for mixed central place systems. In particular, according to the structural dimensions, we can construct the space-filling patterns of fractal central place systems (Figure 5, Figure 6). Suppose that there are two central place systems comprising points, line, and area, which are defined in a 2-dimensional embedding space. The structural dimensions of the two systems are $D_{S1}$ and $D_{S2}$, respectively. If the two systems of central places encounter with each other, then textural dimension of the boundary lines can be calculated using the following formula

$$D_T = D_{S1} + D_{S2} - d, \qquad (9)$$

where $d=2$ denotes the Euclidean dimension of the embedding space. Equation (9) represents one of the important fractal dimension principles, the fractal intersection rule (Vicsek, 1989). If the two central place systems bear the same $k$ value, then $D_S=D_{S1}=D_{S2}$, thus we have

$$D_S = \frac{D_T + d}{2} = 1 + \frac{D_T}{2}, \qquad (10)$$

where $D_S$ denotes the structural fractal dimension of central place systems.

Using the fractal dimension relations, we can calculate the structural dimension of central place systems. For the $k=3$ systems, the textural dimension $D_T=\ln(2)/\ln(3^{1/2})=1.2619$, thus the structural dimension $D_s=1+\ln(2)/\ln(3) = \ln(6)/\ln(3) = 1.6309$. For the $k=4$ systems, the textural dimension $D_T = \ln(3)/\ln(2) =1.5850$, thus the structural dimension $D_s = 1+\ln(3)/\ln(4) = \ln(12)/\ln(4) = 1.7925$. For the $k=7$ systems, the textural dimension $D_T = \ln(3)/\ln(7^{1/2}) = 1.1292$, thus the structural dimension $D_s = 1+\ln(3)/\ln(7) = \ln(21)/\ln(7) = 1.5646$. Based on the concepts of $g$ and $k$ numbers, as indicated above, the textural fractal dimension can be generalized to a common formula as below:

$$D_T = \frac{\ln(g)}{\ln(\sqrt{k})} = \frac{2\ln(g)}{\ln(k)}. \tag{11}$$

Accordingly, in light of equations (10) and (11), the structural fractal dimension can be derived as

$$D_S = 1 + \frac{\ln(g)}{\ln(k)} = \frac{\ln(gk)}{\ln(k)}. \tag{12}$$

Inversely, from equations (11) and (12) it follows equation (10). Using these relations, we can convert the pure textural dimension into pure structural dimension and *vice versa* (Table 2).

Table 2 The fractal dimension relationships between the pure textural fractals and pure structural fractals of central place models

| Type | Textural dimension | Structural dimension | Related fractals |
|---|---|---|---|
| $k=3$ system | $\ln(2)/\ln(3^{1/2})=1.2619$ | $\ln(6)/\ln(3)=1.6309$ | Koch curve and snowflake |
| $k=4$ system | $\ln(3)/\ln(2)=1.5850$ | $\ln(12)/\ln(4)=1.7925$ | Sierpinski gasket and space-filling curve |
| $k=7$ system | $\ln(3)/\ln(7^{1/2})=1.1292$ | $\ln(21)/\ln(7)=1.5646$ | Gosper island |

The above results suggest both spatial and hierarchical intermittency in central places systems. In other words, if we introduce intermittent processes into central place networks and hierarchies, we will obtain fractal central place landscapes. If $g=k$, the central place system has no intermittency, and the dimension will be $D_s=d=2$; If $g<k$, the central place system bears intermittency, and the dimension will be $D_s<d=2$. For the $k=3$ system, $k=4$ system, and $k=7$ system, according to equation (10), the pure structural fractal dimension values are 1.6309, 1.7925, and 1.5646, respectively. Thus we have textural and structural fractal dimension vectors for central place systems as follows

$$\mathbf{V}_T = [1.2619 \quad 1.5850 \quad 1.1292], \quad \mathbf{V}_S = [1.6309 \quad 1.7925 \quad 1.5646],$$

which are based on Christaller's number and only suitable three special central place networks.

Further, the textural dimensions of mixed central place networks can be derived. A mixed central place networks comprise more than one types of pure central place network. For example, a $k=3$ system and a $k=4$ system compose a mixed system of central places. Using equation (9), we can calculate the boundary dimension for any two types of central place systems. For instance, for $k=3$ system and $k=4$ system, the mixed textural dimension of central place boundary is $D_T=D_{S1}+D_{S2}-$ 2=1.6309+1.7925-2=1.4234. The rest may be deduced by analogy (left part in Table 3). Thus we have textural fractal dimension matrix as below:

$$\mathbf{M}_T = \begin{bmatrix} 1.2619 & 1.4234 & 1.1955 \\ 1.4234 & 1.5850 & 1.3571 \\ 1.1955 & 1.3571 & 1.1292 \end{bmatrix},$$

in which the diagonal elements indicate the pure textural fractal dimensions. Then, by arithmetic mean, we can calculate the mixed structural dimensions of different types of central place systems. For example, for $k=3$ system and $k=4$ system, the simple average value of central place networks' structure is $D_T=(D_{S1}+D_{S2})/2=(1.6309+1.7925)/2=1.7117$. The others can be figured out in the similar way (right part in Table 3). Consequently, we have structural fractal dimension matrix as follows

$$\mathbf{M}_S = \begin{bmatrix} 1.6309 & 1.7117 & 1.5978 \\ 1.7117 & 1.7925 & 1.6785 \\ 1.5978 & 1.6785 & 1.5646 \end{bmatrix}.$$

in which the diagonal entries imply the pure structural fractal dimensions. The average dimension of the $k=3$, 4, and 7 systems is 1.6627. The relationships between structural fractal dimension vector and textural fractal dimension matrix can be tabulated as follows (Table 3).

**Table 3 The matrices of mixed textural fractal dimensions and mixed structural dimensions of central place systems**

| Systems/ Dimensions | Textural dimension | | | Structural dimension | | |
|---|---|---|---|---|---|---|
| | $k=3$ system ($D_s$=1.6309) | $k=4$ system ($D_s$=1.7925) | $k=7$ system ($D_s$=1.5646) | $k=3$ system ($D_s$=1.6309) | $k=4$ system ($D_s$=1.7925) | $k=7$ system ($D_s$=1.5646) |
| $k=3$ system ($D_s$=1.6309) | **1.2619** | 1.4234 | 1.1955 | **1.6309** | 1.7117 | 1.5978 |
| $k=4$ system ($D_s$=1.7925) | 1.4234 | **1.5850** | 1.3571 | 1.7117 | **1.7925** | 1.6785 |
| $k=7$ system ($D_s$=1.5646) | 1.1955 | 1.3571 | **1.1292** | 1.5978 | 1.6785 | **1.5646** |

By the fractal dimension values, we can conceive the corresponding spatial structure models of central place systems. In fact, according to the pure structural fractal dimensions, we can construct the space-filling models of fractal central place networks based on intermittency process and patterns (Figure 6). For example, for $k$=4 systems, the fractal landscape comprises self-similar pattern and intermittency process (Figure 7). The others can be deduced by analogy with this process.

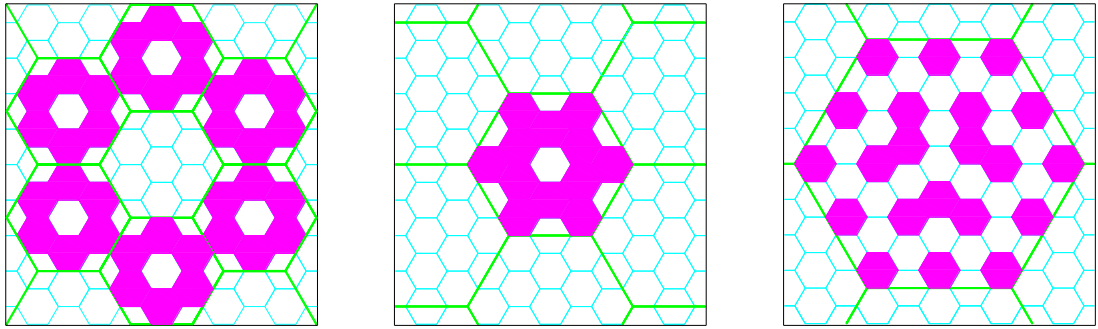

a. $k$=3 system  b. $k$=4 system  c. $k$=7 system

**Figure 6 Sketch maps of intermittency filling patterns of three types of central place systems (the second or third step)**

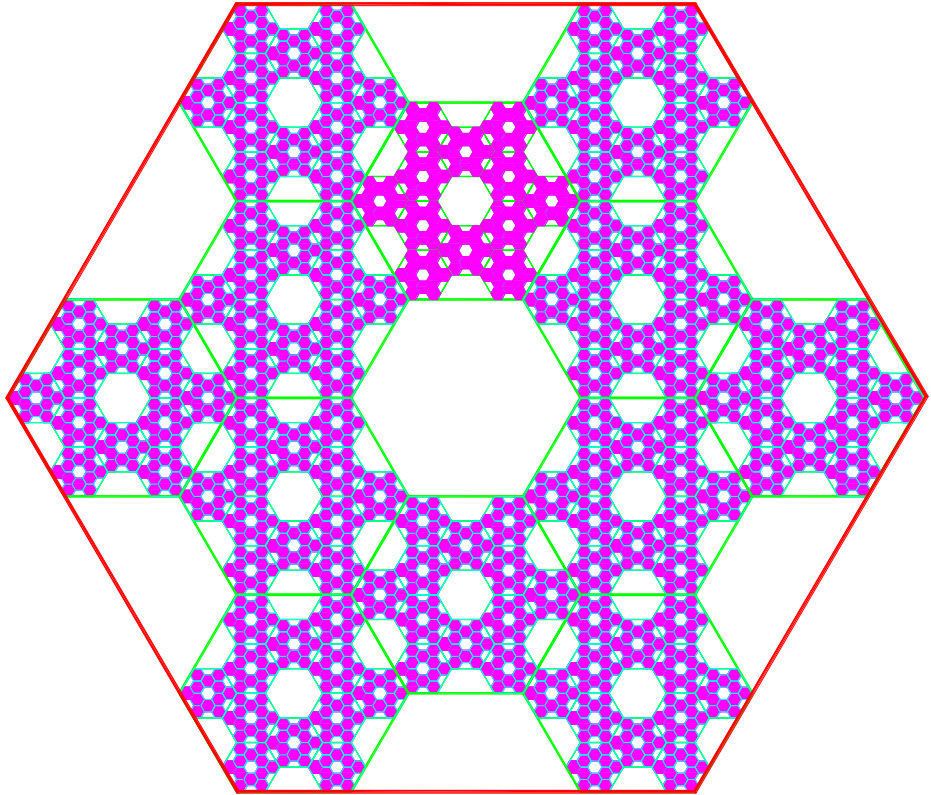

**Figure 7 A sketch map of intermittency filling process and pattern of the $k$=4 central place system (the fourth step)**

# 3. Empirical analysis

## 3.1 Study areas and methods

Cities and urban systems are both complex systems coming between apparent random behaviors and deep spatial orders. The former can be metaphorical as particle behavior, and the latter can be metaphorical as a wave behavior (Chen, 2008). Central place models are the extracted results of hidden orders from human settlement distributions. This theoretical research on central place fractals is based on previous theoretical explorations and empirical analyses. As indicated above, the theoretical base is the central place textural fractals revealed by Arlinghaus and her collaborator (Arlinghaus, 1985; Arlinghaus, 1993; Arlinghaus and Arlinghaus, 1989); the empirical base is the central place structural fractals researched by Chen and his co-workers (Chen, 2011; Chen, 2014; Chen and Zhou, 2006). The observational data from the literature can be employed to make a case study. Seven datasets are available. Four study areas are in southern Germany, namely, Munich, Nuremberg, Stuttgart, and Frankfort, and accordingly, four systems of central places, are taken into account. The datasets were presented by Christaller (1933/1966), and the fractal dimension values are estimated by Chen and Zhou (2006). Two study areas are in America, that is, Iowa and North Dakota, and two systems of central places are considered. The datasets were published by Rayner *et al* (1971), and the fractal dimension values were estimated by Chen (2011). One study area is in China, that is, Henan Province, and a system of urban places is examined. The data were extracted and the fractal dimension was evaluated by Chen (2014).

A central place system is both a self-similar hierarchy and a self-organized network. On the one hand, a central place system bears cascade structure, which can be described with geometric sequences of settlement number, service area, distance, and so on. On the other, a central place system can be treated as an ordered set consisting of nodes and lines. At least five methods can be adopted to calculate the structural fractal dimension of central place systems (Table 4). Three approaches can be utilized to estimate the fractal dimension value of a self-similar hierarchy. The first is the number-distance scaling based on a power law, the second is the characteristic scale ratio based on a pair of exponential laws, and the third is common ratio based on a pair of geometric sequences. At least two approaches can be employed to evaluate the fractal dimension of a self-organized network. The first is also the number-distance scaling, and the second is the well-known box-counting method. Clearly, the number-distance scaling can be applied to both central place hierarchies and networks.

Table 4 Four approaches to estimating the fractal dimension values of central place systems in the real world

| Type | Mathematic model | Method | Equation |
|---|---|---|---|
| **Hierarchy** | Power law | Number-distance scaling | Equation (7) |
| | Exponential laws | Characteristic scale ratio | Equations (3) and (4) |
| | Geometric sequences | Common ratios | Equations (3) and (4) |
| **Network** | Power law | Number-distance scaling | Equation (7) |
| | Power law | Box counting | Equation (13) |

To illuminate the process of fractal dimension estimation, two basic models should be clarified here. One is the number-distance scaling relation, and the other, the boxes' number-size relation. Equation (6) represents an inverse square law and shows a special number-distance relation based on Euclidean space. By introducing the concept of intermittency into central place models, we can use fractal space to replace Euclidean space. Thus equation (6) can be replaced by equation (7), which is the number-distance scaling relation and proved to be equivalent to the boxes' number-size scaling (Chen and Zhou, 2006). The box-counting algorithm is a common approach for estimating fractal dimension values (Beck and Schlögl, 1993). The method have been applied to the studies on fractal cities (Benguigui *et al*, 2000; Chen and Wang, 2013; Feng and Chen, 2010; Shen, 2002; Sun and Southworth, 2013). The box dimension is also termed Minkowski–Bouligand dimension. If the nonempty box number, $N(\varepsilon)$, and the linear size of the boxes, $\varepsilon$, follows a power law such as

$$N(\varepsilon) = N_1 \varepsilon^{-D}, \tag{13}$$

where $N_1$ refers to the proportionality coefficient, and $D$ to the box dimension, then the urban morphology can be regarded as fractal. Equation (13) represents the number-size scaling relation of nonempty boxes in fractal measurement, which can transformed into logarithmic linear form

$$\ln N(\varepsilon) = \ln N_1 - D \ln \varepsilon. \tag{14}$$

Thus the least squares calculation can be used to estimate the fractal parameter $D$. There are two variants of the conventional box-counting method. One is the functional box-counting method (Chen, 2014; Lovejoy, 1987), and the other variant is the grid method (Frankhauser, 1998). All these methods can be treated as generalized box-counting methods for fractal dimension measurement and estimation.

It is important to note that the estimated fractal dimension values of real central place systems differ from the theoretical fractal dimension of central place models. The central place fractal models are all geometric fractals defined in the mathematical world. For regular fractals with an exact self-similar structure, it is possible to evaluate the fractal dimension by a simple theoretical approach (Beck and Schlögl, 1993). In the real world, however, no real fractals can be found. Cities and settlement systems are all random pre-fractals (Addison, 1997), which bear fractal property within certain ranges of scales (Batty and Longley, 1994; Chen, 2008; Frankhauser, 1994). The ranges are termed scaling ranges or scaling region (Wang and Li, 1996; Williams, 1997). In fact, on a double logarithmic scatter plot based on equation (14), the trend line can be divided into three segments. The first segment corresponds to the larger measurement scales, and its slope gives the Euclidean dimension of the fractal's embedding space; the second segment corresponds to the middle measurement scales and represents the scaling range, and is slope gives the estimated value of the fractal dimension; the third segment corresponds to the smaller measurement scales, and its slope gives the topological dimension of the fractal point sets. All the fractal dimension values of the real central places are based on the scaling ranges. We have to employ a proper algorithm to estimate the fractal dimension. Therefore, there are often errors to some extent between the observed values and predicted values of fractal parameters.

## 3.2 Empirical results and findings

The fractal parameter equations can be employed to estimate the fractal dimension values of the central place systems in America, China, and Germany. In fact, the fractal dimensions of the central place system structures were evaluated in previous works. First, the structural fractal dimensions of the central place system in Munich, Nuremberg, Stuttgart, and Frankfort of Germany are estimated with both characteristic scale ratio and number-distance scaling (Chen and Zhou, 2006); second, the structural fractal dimensions of central place systems in Iowa and North Dakota of USA were estimated with the number-distance scaling (Chen, 2012). Third, the structural fractal dimension of the central place system in Henan Province of China was evaluated with the box-counting method (Chen, 2014). In the previous papers, the fractal dimension values of the central place systems in southern Germany are mainly based on the characteristic scale ratio method. In this paper, the fractal dimensions are recalculated using the number-distance scaling method (Table 5). The two sets of calculation results are close to one another. That is, there are no significant differences between the fractal dimension values based on two exponential laws and those based on a power law.

Table 5 The fractal models and structural fractal dimension values of seven systems of urban places in the real world

| Country | Region | Models | Fractal dimension $D_S$ | Goodness of fit $R^2$ |
|---|---|---|---|---|
| Germany | Munich | $f_m = 2631.2477 L_m^{-1.7327}$ | 1.7327 | 0.9614 |
| | Nuremberg | $f_m = 3004.4598 L_m^{-1.6852}$ | 1.6852 | 0.9810 |
| | Stuttgart | $f_m = 5657.0290 L_m^{-1.8370}$ | 1.8370 | 0.9758 |
| | Frankfort | $f_m = 1088.4088 L_m^{-1.4811}$ | 1.4811 | 0.9782 |
| America | Iowa | $f_m = 12708.1052 L_m^{-1.8763}$ | 1.8763 | 0.9564 |
| | North Dakota | $f_m = 9878.1356 L_m^{-1.7555}$ | 1.7555 | 0.9799 |
| China | Henan | $N(\varepsilon) = 1.1013 \varepsilon^{-1.8585}$ | 1.8585 | 0.9948 |

**Sources:** Chen (2012), Chen (2014), and Chen and Zhou (2006). **Note:** The fractal dimension value of Henan's urban system was evaluated by the box-counting method, differing from those of American and German systems of central places which were estimated by the number-distance scaling.

As indicated above, the textural fractal dimension of real central place systems cannot be directly estimated by means of empirical methods. All the above-shown fractal dimension values are actually the structural fractal dimensions. Using the fractal dimension equations proposed in this paper, we can estimate the textural fractal dimensions through the structural fractal dimension values. By means of equation (10), the structural dimension can be easily converted into the textural dimension (Table 6). The formula is $D_T = 2(D_S - 1)$. Sometimes the results are abnormal. The textural dimension values should come between 1 and 2. However, the parameter of Frankfort's system is less than 1, and this is not acceptable. According to the theoretical inference, its textural dimension value can be regarded as 1. Comparing the empirical fractal dimension values with the theoretical values displayed in Table 3, we can bright to light the dominated principles behind a central place system. For example, Frankfort's system is dominated by the separation principle, Stuttgart's, Iowa's, and Henan's systems are dominated by the traffic principle, and Munich's, Nuremberg's and North Dakota's systems may be dominated by both market principle and traffic principle. Several conclusions can be reached as follows. First, the most important principle is traffic principle, which maybe play a leading role in the evolution process of central place systems. Second, the market principle is always associated with the traffic principle. Third, a few central place systems are dominated by the separation principle. Where the central place systems in southern Germany are concerned, the domination principle of a system predicted by fractal dimensions often differs from the empirical judgment of Christaller (1966). The discovery of the status of the traffic principle is valuable in future urban studies. Anyway, according to Batty (2013), a city is not simply a place in space, as understood by traditional ideas; a city a system of networks and flows. To understand geographical space, we must understand flows, networks, and the relations between elements that compose the city system. By analogy, a central place system is not a set of urban places in regional space, but a complex network of human settlements. The spatial interaction measured by flows and influenced by transport principle is the key to understanding central place development.

**Table 6 The fractal dimension values and the corresponding principles of seven systems of urban places in the real world**

| Country | Region | Fractal dimension $D_S$ | Fractal dimension $D_T$ | The $k$ value | Principle |
|---|---|---|---|---|---|
| **Germany** | Munich | 1.7327 | 1.4654 | 3 and 4 | Market and traffic |
|  | Nuremberg | 1.6852 | 1.3704 | 3 and 4 | Market and traffic |
|  | Stuttgart | 1.8370 | 1.6740 | 4 | Traffic |
|  | Frankfort | 1.4811 | (0.9622) | 7 | Separation |
| **America** | Iowa | 1.8763 | 1.7526 | 4 | Traffic |
|  | North Dakota | 1.7555 | 1.5110 | 3 or 4 | Market and traffic |
| **China** | Henan | 1.8585 | 1.7170 | 4 | Traffic |

**Note**: The textural fractal dimension value of Frankfort's system of central place can be treated as 1.

## 4. Discussion

The advantage of fractal geometry is to describe the complex systems without characteristic scales. Cities and systems of cities bear scaling invariance indicating spatial complexity, which can be characterized by fractal parameters (Batty, 2005; Chen, 2008). Central place theory represents the basic models for urban systems. The relationships between fractal texture and fractal structure are important for us to understand central place systems in the real world. The structural fractal dimensions can be associated with the textural fractal dimension by a mathematical formula. The analytical process of this study is as follows. **First, revealing fractal patterns of central place models.** Three regular fractals are brought to light from central place models, including Koch snowflake, Sierpinski space-filling curve, and Gosper island. These fractals indicate central place texture. **Second, revealing fractal dimension relations.** By means of the well-known fractal dimension rules, we can derive pure structural fractal dimension values from the pure textural fractal dimension values of central place models ($D_s$=1.6309, 1.7925, and 1.5646 for $k$=3, $k$=4, and $k$=7 systems). Further, from the pure structural fractal dimension values, we can derive mixed textural fractal dimension values and calculated average structural fractal dimensions. **Third, formulating central place fractal dimensions.** The $g$ number is defined by the element number of fractal generators. Combining the g number with the $k$ number, we can express the textural fractal dimensions and structural fractal dimensions with two formulae, i.e., equations (11) and (12). The formulae can be used to explain or predict the central place fractals based on intermittency. **Fourth, modeling fractal patterns of central place system structure.** In terms of the structural fractal dimension values, new fractal models can be constructed for the $k$=3, $k$=4, and $k$=7 central place systems (Figures 6 and 7).

The fractal properties of central place systems have been studied for a long time, and a number of interesting findings have been published. Compared with the previous works on central place fractals in literature (Arlinghaus, 1985; Arlinghaus and Arlinghaus, 1989; Batty and Longley, 1994; Chen, 1998a; Chen, 1998b; Chen, 2012; Chen and Zhou, 2006; Frankhauser, 1998; Frankhauser, 2008), progress has been made as below (Table 7). First, new fractals such as Koch snowflakes and Sierpinski space-filling curve are revealed. The results are illustrated by Figures 4 and 5. In previous literature, only Koch snowflake in the $k$=3 system is revealed. In this work, the Koch snowflakes in the $k=3$, $k=4$, and $k=7$ systems are all brought to light. Second, new structural central place fractal models of the $k=3$, $k=4$, and $k=7$ systems are constructed. The results are illustrated by Figures 6 and 7, which are useful for us to understand the structure of the real central place systems. Third, the transformation relation between textural fractal dimension and structural fractal dimension in central place models are formulated. The results are expressed by equations (10), (11), and (12). These findings are helpful for us to understand the evolutional process and patterns of urban and rural settlement systems in the real world.

Table 7 The main new points of this paper manifested by comparison with the previous works

| Literature | Main findings |
|---|---|
| **Arlinghaus (1985)** | Fractal boundaries and textural fractal dimensions of Christaller's central place models |
| **Arlinghaus and Arlinghaus (1989)** | Fractal boundaries and textural fractal dimensions of Lösch's central place models |
| **Chen (1998a)** | Empirical evidences of the Koch snowflake model of the $k=3$ central place system |
| **Chen (1998b)** | Random Sierpinski network model of the $k=4$ central place systems |
| **Chen and Zhou (2006)** | Theoretical fractal models and empirical evidences of Christaller's central place systems in southern Germany |
| **Chen (2011)** | Theoretical models and empirical evidence of Christaller's central place fractal hierarchies |
| **Chen (2014)** | Multifractal models and description of central place systems |
| **This paper** | (1) Mathematical relation between textural and structural fractal dimensions of central place models; (2) Spatial structure models of Christaller's central place systems; (3) Koch snowflake curves in $k=3,4$, and 7 systems and mixed Sierpinksi space-filling curve in the $k=3$ system |

The main shortcomings of these studies lie in three aspects. First, the average fractal dimension of central place structure is not satisfying. According to the union rule, the union of two fractal sets A and B with fractal dimensions $D_A>D_B$ has the dimension $D=D_A$ (Vicsek, 1989). In other words, the fractal dimension of two mixed central place systems A and B is $D= \max (D_A, D_B)$ rather than $D=(D_A+D_B)/2$. I'm not sure whether the mixed central place system follow the fractal union rule. Second, the models have not been associated with systematic empirical analysis. The value of a theoretical model depends on its effect of explaining or predicting reality. However, it is difficult to obtain a complete dataset for central place fractal studies. Third, the simulation experiments have not been made for these models. Mathematical modeling is based on macroscopic level of a theoretical law, while an experiment can be employed to reveal the microscopic mechanism of system evolution. Laboratory experiment cannot be used to make urban studies. If and only if the theoretical modeling, empirical study, and computer simulation are successfully integrated into a logic framework on urban places or human settlements, the exploration process on central place fractals will be really fulfilled. Further work remains to be done in the future.

## 5. Conclusions

Central place theory is a classical doctrine in human geography, but central place fractals cause many new problems. This work is devoted to pure theoretical exploration. The results and findings are revealing for our understanding the self-organized evolution of human settlements. Using the theoretical models, maybe we can develop useful spatial optimization methods. From the analytical process and results, the main conclusions are drawn as follows. **First, the textural fractal dimension of a central place system can be linked to its structural fractal dimension by a linear equation.** The textural fractals and the structural fractals are complementary. It is easy to obtain the fractal dimension values of the central place fractal models in the mathematical world, and it is convenient to estimate the structural fractal dimension of a central place system in the real world. However, it is difficult to calculate the textural fractal dimension of a real central place system. Using the fractal dimension equation, we can estimate the textural dimensions through the structural dimensions. **Second, two numbers can be employed to formulate the textural fractal dimensions and structural fractal dimensions of central place systems**. The two numbers, $k$ and $g$, represent the common ratio of the geometric sequences of a central place hierarchy and the amount of fractal units in a fractal generator, respectively. Using the two numbers, we can reformulate the fractal dimension of central place models. The features of the spatial structure of different central place models can be characterized by the two numbers. **Third, the traffic principle play a leading role in central place system evolution.** This differs from the traditional viewpoint that the marketing principle play the more important role in central place development. Fractal dimension values imply space filling extent. The traffic principle results in higher space filling, while the separation principle leads to the lower space filling. The space filling degree resulting from marketing principle is in the middle. According to the empirical evidences, the fractal dimension values of the real central place systems are mainly consistent with the central place model based on the traffic principle rather that based on the marketing principle.

## Acknowledgements

This research was sponsored by the National Natural Science Foundation of China (Grant No. 41671167). The support is gratefully acknowledged.

# Appendices

## Appendix 1 Fractal takes central places

The structural fractals of central place models can be derived from the textural fractals. The central place fractals were brought to light by Arlinghaus (1985, 1993), who illustrated the initiators and generators of fractals for $k$=3, 4, and 7 systems clearly. Alternatively, the generation of central place fractals can be sketched by Figures A1 and A2, in which the construction process of the $k$=4 system is different from but equivalent to the one proposed by Arlinghaus (1985).

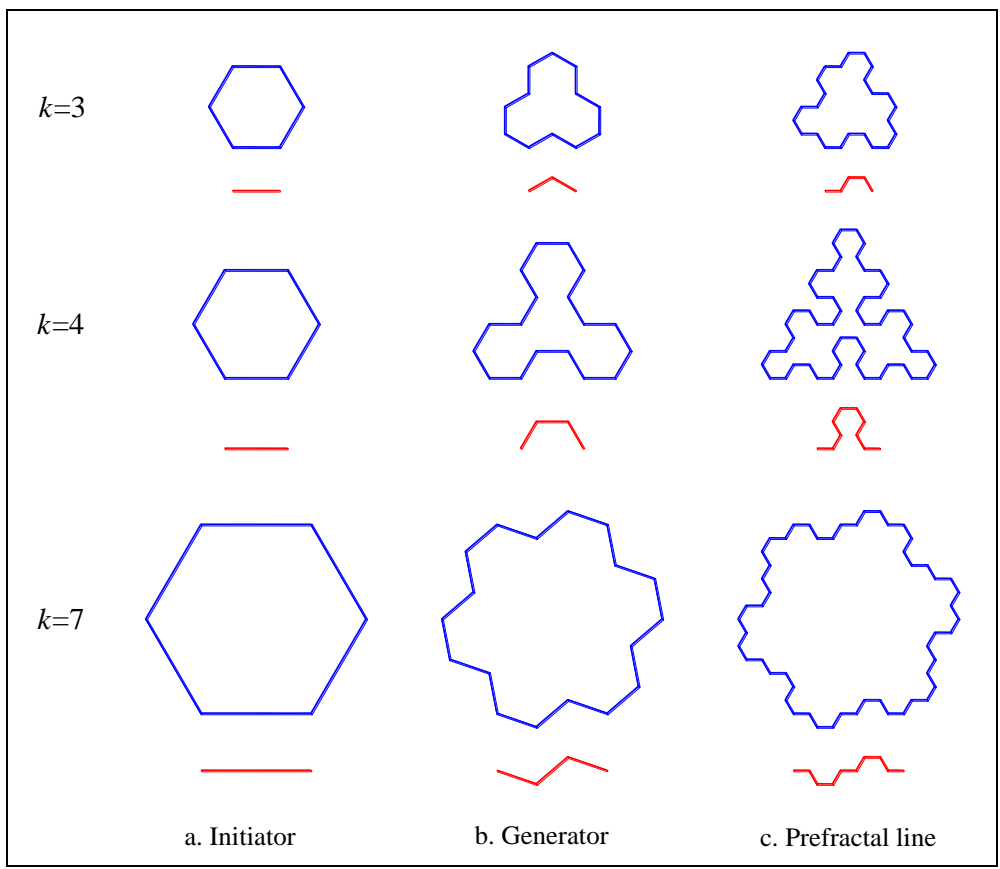

**Figure A1 The initiators, generators, and boundary lines of three types of central place fractals (the first three steps). Note**: The generator of the $k$=4 system differs from one proposed by Arlinghaus (1985).

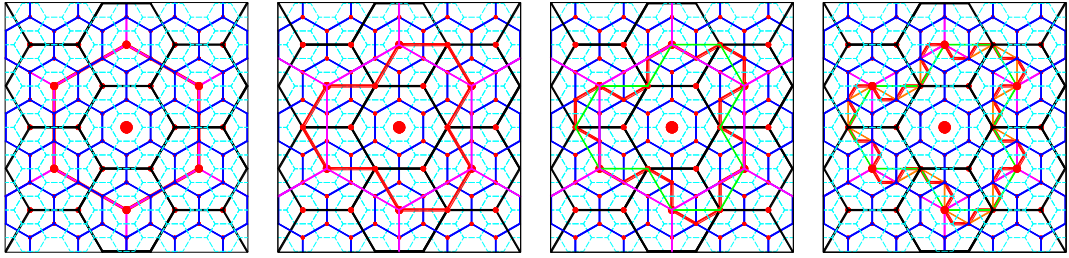

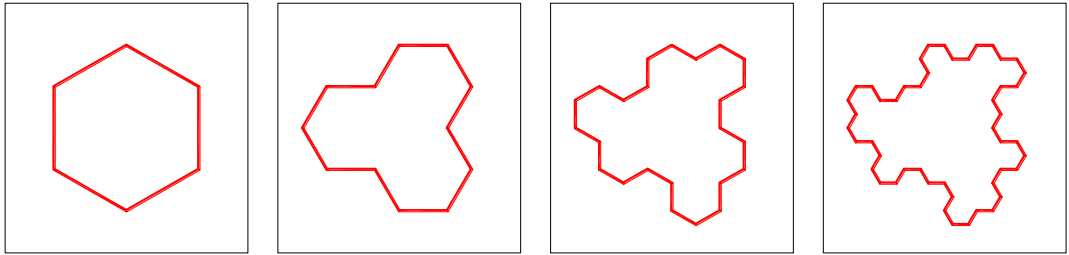

a. Fractal boundary for *k*=3 systems

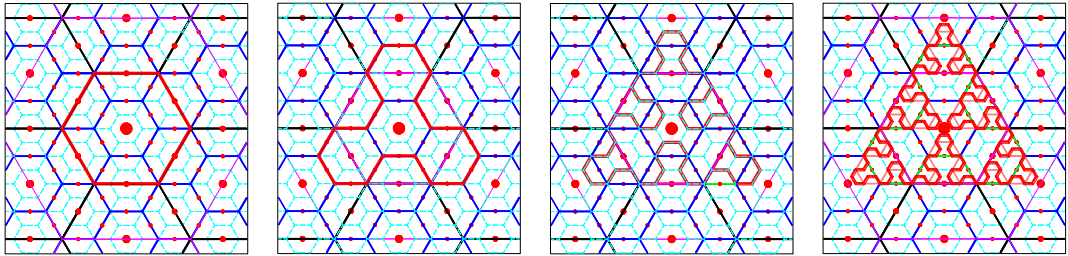

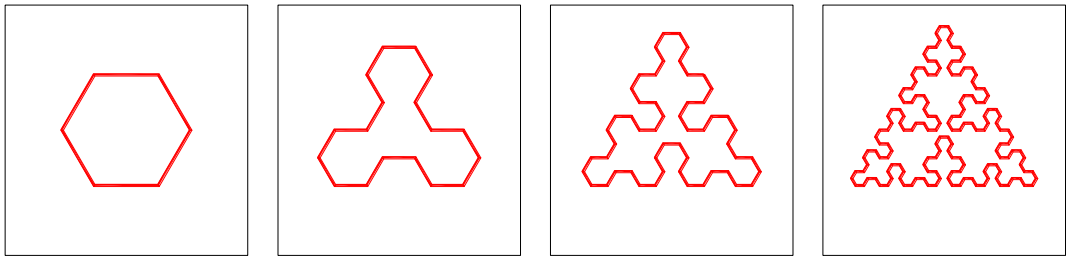

b. Fractal boundary for *k*=4 systems

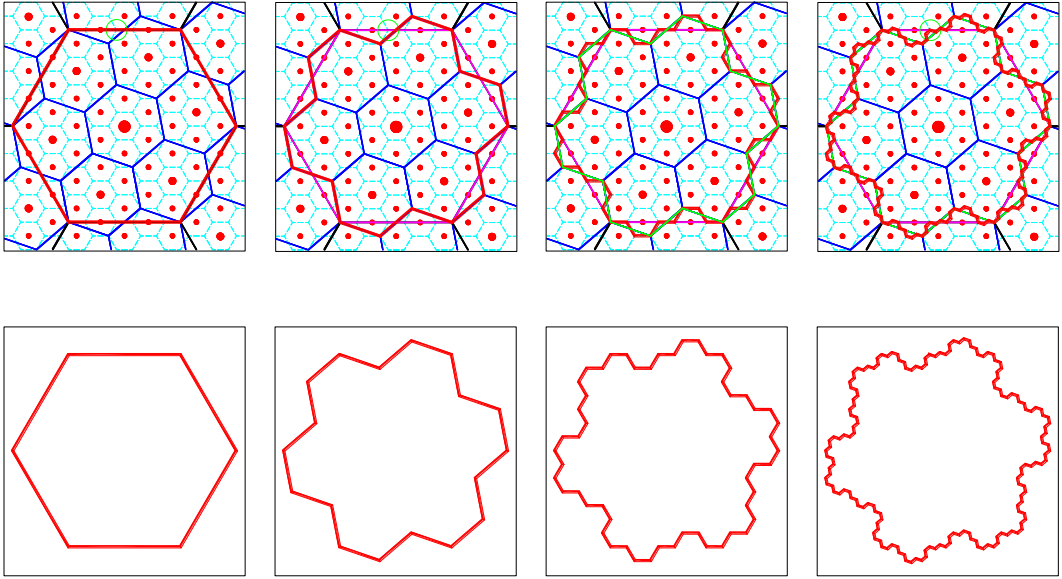

c. Fractal boundary for *k*=7 systems

**Figure A2 The generation processes of three types of central place boundaries (the first four steps) Note**: The mode of the *k*=4 system differs from the pattern presented by Arlinghaus (1985). A hierarchy of regional units forms a textural pattern of central place fractals.

## Appendix 2 Two variants of Sierpinski gasket

Three types of Sierpinski gaskets are involved in this study. The first one is classical Sierpinski gasket. In its generator, there are 3 elements ($N$=3), and the linear size of these elements is $r$=1/2. Thus, the fractal dimension is $D$=ln(3)/ln(2)=1.5850 (Figure A3(a)). The second one is extended Sierpinski gasket. Its generator includes 6 elements ($N$=6) with linear size $r$=1/3. This suggests that the fractal dimension is $D$=ln(6)/ln(3)=1.6309 (Figure A3(b)). The third one is incorporated Sierpinski gasket, which is in fact the interlaced pattern of three classical Sierpinski gaskets (Figure A3(c)).

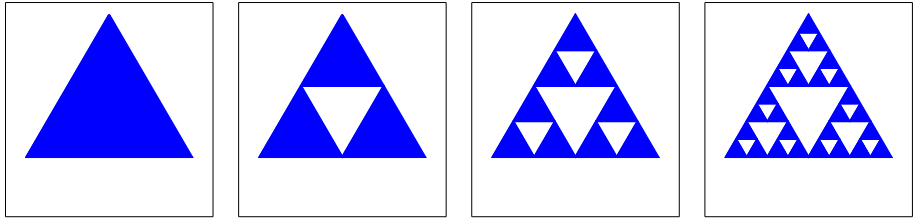
a. Sierpinski gasket ($D=\ln(3)/\ln(2)=1.585$)

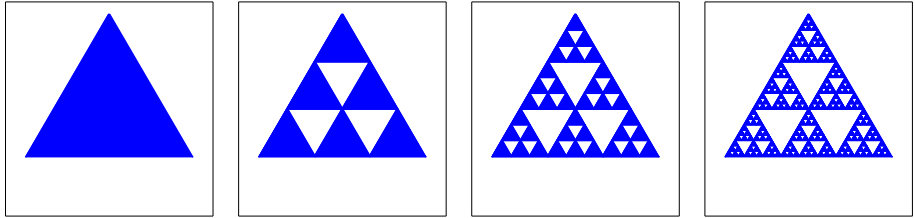
b. Extended Sierpinski gasket ($D=\ln(6)/\ln(3)=1.631$)

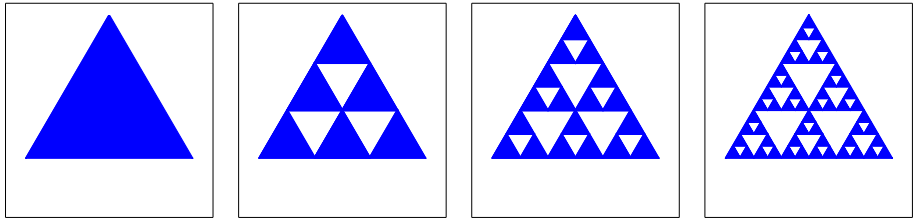
c. Incorporated Sierpinski gasket ($D=\ln(3)/\ln(2)=1.585$)

**Figure A3 The standard Sierpinski gasket, generalized Sierpinski gasket, and interlaced Sierpinski gasket (the first four steps)**